# Breaking Bad: Detecting malicious domains using word segmentation

Wei Wang
AT&T Security Research Center
New York, NY 10007
Email: wei.wang.2@att.com

Kenneth E. Shirley
AT&T Labs Research
New York, NY 10007
Email: kshirley@research.att.com

*Abstract*

In recent years, vulnerable hosts and maliciously registered domains have been frequently involved in mobile attacks. In this paper, we explore the feasibility of detecting malicious domains visited on a cellular network based solely on lexical characteristics of the domain names. In addition to using traditional quantitative features of domain names, we also use a word segmentation algorithm to segment the domain names into individual words to greatly expand the size of the feature set. Experiments on a sample of real-world data from a large cellular network show that using word segmentation improves our ability to detect malicious domains relative to approaches without segmentation, as measured by misclassification rates and areas under the ROC curve. Furthermore, the results are interpretable, allowing one to discover (with little supervision or tuning required) which words are used most often to attract users to malicious domains. Such a lightweight approach could be performed in near-real time when a device attempts to visit a domain. This approach can complement (rather than substitute) other more expensive and time-consuming approaches to similar problems that use richer feature sets.

*Keywords*-Mobile Network Security, Domain Name Analysis, Text Mining, Logistic Regression, Word Segmentation, Lasso Regularization

## I. INTRODUCTION

The frequency of mobile attacks has increased dramatically in recent years with the ubiquitousness of smartphones [1], [2]. Numerous mobile attacks have utilized poorly managed hosting sites or newly registered domains to serve as phishing sites, command-and-control (aka C&C) servers or to host malware binaries [3]–[5]. It is of great interest to users and network service providers to be able to detect which domains are malicious as quickly as possible and to proactively protect users from visiting them.

Previous approaches to detecting malicious domains and URLs have used machine learning algorithms trained on historical data to measure the extent to which various features of these domains or URLs are correlated with their "maliciousness", which is often measured by a reputation score or membership in a manually verified blacklist. There is a general tradeoff between speed and accuracy in these approaches: Greater accuracy is often only possible by using computationally intensive feature generation methods, such as retrieving web content or using metadata from DNS lookups and WHOIS queries. Our approach differs from these in that we only use lexical features of domain names in our algorithms, providing a fast, lightweight solution.

Note that by only considering domain names in our detection algorithm, rather than full URLs, the set of potential lexical features is of limited size. Unlike full URLs, which can be split into tokens, for example, by delimiters such as slashes and periods, domain names, by definition, can only contain alphanumeric characters and hyphens. To overcome this challenge, we use a word segmentation algorithm to segment domain names into individual tokens, generating large numbers of new features with which to train our classification models. We show that using token-based features in our models provides substantial increases in predictive accuracy relative to models that only use traditional quantitative lexical features of domain names, such as length, the number of digits, and Markov model-based features, among others. Furthermore, we can also interpret our fitted models to learn which words attract users to malicious sites, and which words are typically associated with benign domains. Our experiments are performed on domain names that were visited on a cellular network, although our methodology is generalizable to domain names visited on a wired network.

## II. RELATED WORK

Several researchers have studied the detection of malicious domains and URLs using machine learning in recent years. In a relatively early paper, Gerara et al. [6] used logistic regression on 18 features with approximately 2500 training examples to model membership on a blacklist of URLs. They included features in which tokens were obtained using an algorithm to extract long, common substrings of URLs, which identified 150 words across the roughly 1200 positive examples, and they subsequently manually trimmed this word list to the eight words that they considered the most likely to have predictive power. McGrath and Gupta [7] studied characteristics of phishing URLs and domains, discovering that various features such as domain name length, frequencies of certain characters, and the presence of certain brand names in URLs were associated with phishing, but they limited themselves to a descriptive study, rather than developing a predictive model. He et al. [8] used several Markov models trained on different text corpora to create features which captured the fact that legitimate domains often contain meaningful English words while many malicious domain names do not. They also extracted lexical features in new domain names as well as features in DNS data, and used various machine learning methods to classify new domain names. Bilge et al. [9] used a set of 15 DNS-related features to predict domain name maliciousness, two of which were lexical characteristics of the domain names: the percentage of characters that were numerical, and the proportion of the domain name that was comprised by the longest English-language word; they did not, however, use individual words as features. Yadav et al. study algorithmically generated malicious domain names [10] by looking at the distributions of unigrams and bigrams in domain names that are mapped to the same set of IP addresses. They compare the performance of several distance metrics, including KL-distance, edit distance and Jaccard measurement to identify new domains that are suspiciously similar (measured via distance) to known bad domains. Last, Ma et al. [11] used machine learning methods to detect malicious URLs in a paper that is the most relevant to our work. They used both lexical and host-based features of URLs in several supervised learning algorithms to differentiate benign and malicious URLs. They include words as features, but they limit their vocabulary of words to those that are separated by a specific set of punctuation marks in the domain name or URL path. None of these machine learning approaches to malicious domain detection have used word segmentation on the domain names to generate thousands of additional features as we do.

Word segmentation, defined here as transforming a single string into a sequence of one or more non-empty substrings, has been applied to domain names in previous work. Wang et al. [12] perform a series of experiments performing word segmentation on URLs. They manually annotate each URL in their experiment with a correct segmentation (including multiple plausible options if applicable) and they measure which training corpus (from a small set of candidates) maximizes their algorithm's performance recovering the known, true segmentations in their labeled data. Srinivasan et al. [13] extend this approach by incorporating the lengths of segments into their algorithm, and they evaluate their performance on domain names that have been manually segmented by human annotators. In contrast to these approaches, rather than evaluating word segmentation by its ability to recover a known correct segmentation, we focus on how well word segmentation works in the context of predicting malicious domains.

## III. DATA

Our domain name data consists of what are known as *effective second-level domain names*. Given a full URL, we first identify the top-level domain (TLD) according to the public suffix list [14]. Then, the effective second-level domain name is defined as the combination of the TLD and the string of characters between the TLD and the second period to the left of the TLD. For example, in the URL `http://www.more.example.com/path-to-url.html`, the TLD is `com` and the effective second-level domain name is `example.com`. For the rest of the paper, a "domain" or a "domain name" refers to the effective second-level domain name.

The domains that we used in our experiments came from two sources. First, we gathered a sample of "fresh" domains that were visited on the cellular network of a large U.S. cellular provider during the month of September 2014. A "fresh" domain for a given day was defined as any domain that was visited that day, but that had not been visited during the preceding 30 days (on the cellular network). In other words, these domains consist of a mixture of brand-new domains (that, perhaps, came into existence during or shortly before September 2014) and relatively unpopular domains (that may have been in existence for a long time prior to September 2014, but had not been visited recently). From 26 days in September (omitting 4 due to data feed interruptions), we gathered a sample of 1,372,120 fresh domains ($> 52,000$ per day), all of which were, by definition, unique. Second, we gathered a random sample of 30,000 domains from DMOZ, the open directory project [15] on November 20, 2014 (out of 2,472,408 unique fully qualified domain names listed there at the time).

The outcome variable that we trained our models to predict is based on the Web of Trust (WoT) reputation rating [16] (which was also used by [10]). The WoT reputation rating is an integer in $\{0, 1, ..., 100\}$ that describes how trustworthy a domain is, according to a combination of user-generated reviews as well as a domain's membership on various public blacklists and other third-party systems. WoT also provides a confidence score for each rating, which is also an integer in $\{0, 1, ..., 100\}$ denoting how reliable the reputation rating is for a given domain, where higher confidence scores indicate more reliable ratings. Furthermore, WoT reputation ratings come in two categories: "Trustworthiness" and "Child Safety". We only consider the "Trustworthiness" rating in our experiments.

The WoT API documentation [17] explains that a WoT reputation score, denoted $r$, falls into one of five labeled categories:

1) $0 \leq r < 20$: Very Poor
2) $20 \leq r < 40$: Poor
3) $40 \leq r < 60$: Unsatisfactory
4) $60 \leq r < 80$: Good
5) $80 \leq r \leq 100$: Excellent

The WoT API documentation also suggests a threshold on the confidence score $c$ such that a rating is reliable only if $c \geq 10$, but further suggests that different thresholds may be optimal for different applications.

In our experiments, we defined the binary variable "maliciousness", denoted $m_i$, based on the WoT rating for domain $i$, denoted $r_i$, where the domain is "malicious" ($m_i = 1$) if $r_i < 60$, and the domain is "safe" ($m_i = 0$) otherwise.

We queried the WoT API for reputation ratings (and confidence scores) for all 1,372,120 domains in our sample of cellular network domains, and we successfully gathered a reputation rating and confidence score for the "Trustworthiness" category for 261,623 of them. These queries were performed approximately 40-60 days after the dates on which the domains were visited in the cellular network (such that for new domains, there was sufficient time for users to flag the domain as malicious, but not so much time that a high proportion of malicious domains would revert to being benign before we queried their ratings). Following previous work [11], we assumed that all the DMOZ domains were benign ($m_i = 0$), since they are curated by editors. A few summary statistics of these WoT reputation ratings and confidence scores are listed in Table I.

TABLE I
SUMMARY OF WOT DATA FOR CELLULAR NETWORK DOMAINS

| Statistic | Reputation Ratings | Confidence Scores |
|---|---|---|
| Mean | 61.1 | 7.2 |
| Median | 62 | 6 |
| St. Dev. | 18.0 | 6.3 |
| # < 60 | 38,377 | - |
| # ≥ 60 | 223,246 | - |
| # < 10 | - | 180,191 |
| # ≥ 10 | - | 81,432 |

We ran three experiments to measure how accurately we could predict the maliciousness of domain names based on different feature sets. The three experiments are differentiated by which subsets of data formed the training and test sets:

1) The "balanced" data set consisted of 30,000 randomly sampled malicious domains visited on the cellular network (out of 38,377), and the 30,000 randomly sampled safe domains from DMOZ. 15,000 from each group were randomly sampled to form the training set, and the other 15,000 of each group were designated as the test set. The baseline rate of maliciousness in this data set is (30,000)/(60,000) = 50%.
2) The "unfiltered cellular" data set consisted of all 261,623 domains visited on the cellular network, with 80% randomly sampled as the training set, and the remaining held out as the test set. The baseline rate of maliciousness in this data set is (38,377)/(261,623) = 14.7%.
3) The "filtered cellular" data set consisted of a subset of 80,077 domains visited on the cellular network (out of 261,223) that had a confidence score $c \geq 10$ and a reputation rating $r \in \{0, 1, ..., 39\} \cup \{60, 61, ..., 100\}$, omitting low-confidence ratings and ratings in the middle of the scale (which might indicate that the maliciousness of the domain is ambiguous). The baseline rate of maliciousness in this data set is (19,639)/(80,077) = 24.5%.

## IV. FEATURES

The features that we generated to predict maliciousness fall into five groups: (1) Basic features, (2) Character indicator variables, (3) Log-likelihood, (4) Top-level domains, and (5) Words. Each set of features was measured for each of the 261,623 (cellular) + 30,000 (DMOZ) = 291,623 domains in our data.

### A. Basic Features

We measured four "basic" features:
1) **number of characters:** the number of characters in the domain name, excluding the top-level domain and all periods. (mean = 11.5).
2) **number of hyphens:** the number of hyphens in the domain name (mean = 0.12).
3) **number of digits:** the number of digits in the domain name (mean = 0.09).
4) **number of numbers:** the number of *numbers* in the domain name, where a *number* is defined as a string of consecutive digits of length > 0 (mean = 0.04).

To allow for non-linear relationships between these features and the outcome, we binned the number of characters into deciles, and the other three basic features into the bins {0, 1, 2, ≥ 3}. The hypothetical football-related domain "4downs-10yards.com", for example, contains 14 characters, 3 digits, one hyphen, and two numbers. Rather than 4 feature vectors, the binned versions of these four basic features contain 10, 4, 4, and 4 feature vectors, respectively.

### B. Character Indicator Variables

We created 36 binary features to measure the presence of each character from "a" to "z" and each digit from 0 to 9 in each domain name. The most and least frequently occurring characters were "e" and "q", respectively (occurring at least once in 198,304 and 5,152 domain names, respectively), and the most and least frequently occurring digits were "1" and "7", respectively (occurring at least once in 3,980 and 1,193 domain names, respectively).

### C. Log-likelihood

We computed the log-likelihood of the sequence of characters in each domain name using a first-order Markov model as the probability model for characters in the English language. To define the transition probabilities between characters, we computed the table of first-order transitions from a list of the top 1/3 million unigrams from the Google Ngrams corpus [18]. For each domain, we removed digits and hyphens, and then computed the log-likelihood of the sequence of remaining characters, using the distribution of first characters in the unigram list as the probability distribution of first characters for each domain. We also computed a normalized version of this feature in which we divided the log-likelihood by the number of characters in the domain (omitting digits and hyphens) to account for the fact that longer domains on average have lower log-likelihoods. Last, we binned these values into deciles (as we did with the "number of characters" feature), and we included an additional, 11th bin for the 197 domain names in the data that only contained digits and hyphens (and thus had a missing value for the log-likelihood).

### D. Top-level domains

We observed 857 unique top-level domains (TLDs) in our data, where each domain belongs to exactly one TLD (defined as existing on the Mozilla Public Suffix List [14]). The most common TLD is ".com" (58% of domains in our sample), and the next 5 most common are ".org", ".net", ".de", and ".co.uk" (11%, 5%, 3%, and 3% of domains, respectively). 337 of the 857 TLDs were only observed once.

*E. Words*

Another source of features in domain names are individual words. First, we define some notation. In a set of domain names, we refer to a *token* as a single occurrence of a *word* within a domain name, where each unique token type in the data defines a *word*. The *vocabulary* is the list of all words observed across the collection of domain names. If the domain "duckduckgo.com", for example, were segmented into three tokens, {"duck", "duck", "go"}, it would only contain two words: "duck" and "go".

Since all domain names in our data are unique, it would be impossible to statistically learn anything about maliciousness by treating the entire domain names themselves as tokens. Domain names often, however, contain occurrences of words concatenated together, which are often human-readable, attracting users to websites. We attempt to extract individual word occurrences (i.e. tokens) from our set of domain names, allowing us to learn the associations between individual words and maliciousness in our training data, which, in turn, allows us to predict maliciousness of new domains that contain individual words in common with those in our training set.

To do this, we use a word segmentation algorithm (here "word" is used in the informal sense) described by Norvig [19]. Using a sample from the the Google bigrams corpus to define a probability model for pairs of consecutive words (including a default low probability for pairs involving a word that does not exist in the bigram list), Norvig describes a dynamic programming algorithm to compute the most likely segmentation of a string of characters into a set of one or more tokens. We applied this algorithm to the 291,623 domain names in our data by ignoring periods and the TLDs, and then applying the segmentation algorithm to each hyphen-separated substring of each domain name. In essence, we treat the hyphens as known token boundaries, and then we segment the resulting substrings using Norvig's algorithm to extract additional tokens that are not hyphen-separated. The result is a list of 682,253 tokens observed across the 291,623 domain names, for an average of 2.34 tokens per domain (with a range of 1 to 11 and a mode of 2). This average is quite close to the 2.24 and 2.21 average tokens per domain reported in two experiments in [13], although less than the 2.66 average tokens per URL from [12], which is reasonable since domain names are substrings of URLs. The 30 most common words and their frequencies are listed in Table II. Although many of the most frequent words, such as "a", "of", and "the", would be discarded as "stop words" in other text mining applications, we included them in our feature set because our modeling method performs regularization such that features with no signal will be ignored (i.e. their coefficients will be set to zero) "automatically". Overall there were 94,050 words in the vocabulary. Rather than using word frequencies within domains as features, we simply recorded the 0/1 presence of a word in a domain in our feature construction so that the scale of these features matched those of the other feature sets (and because it didn't make a large difference: only 0.3% of domain names contained more than one occurrence of a given word).

|    | Word   | Freq |    | Word | Freq |    | Word   | Freq |
|----|--------|------|----|------|------|----|--------|------|
| 1  | a      | 6772 | 11 | on   | 1638 | 21 | shop   | 1168 |
| 2  | the    | 6186 | 12 | my   | 1593 | 22 | world  | 1157 |
| 3  | i      | 3515 | 13 | is   | 1426 | 23 | music  | 1143 |
| 4  | of     | 3183 | 14 | club | 1393 | 24 | city   | 1078 |
| 5  | s      | 2789 | 15 | web  | 1380 | 25 | it     | 1056 |
| 6  | in     | 2630 | 16 | art  | 1297 | 26 | center | 1047 |
| 7  | online | 2554 | 17 | inc  | 1234 | 27 | news   | 1034 |
| 8  | and    | 2495 | 18 | co   | 1187 | 28 | st     | 1032 |
| 9  | e      | 1847 | 19 | for  | 1181 | 29 | free   | 1017 |
| 10 | to     | 1772 | 20 | de   | 1171 | 30 | group  | 1012 |

TABLE II
TERM FREQUENCIES

V. MODELS AND METHODS

For each of our three experiments and each of our seven combinations of feature sets (described in the following paragraph), we fit a logistic regression model using the lasso penalty [20] to regularize the parameter estimates. The lasso places a penalty on the $L_1$ norm of the regression coefficients, and has the effect of setting many coefficients to zero, which aids interpretability as well as improves out-of-sample predictions by avoiding overfitting. Ma et al [11] found that lasso-regularized logistic regression models performed as well in a similar task to other methods such as Naive Bayes and support vector machines. In our experiments we chose the lasso penalty parameter by using 10-fold cross-validation on the training data, and choosing the largest value of the penalty parameter that gave a cross-validated AUC within one standard error of the minimum [20] (this is known as the "one standard error rule"). We used the R package `glmnet` [21].

We fit seven models (named M1 - M7) to our training data in each of the three experiments. Each of the seven models is defined by its use of a different combination of the five sets of features. The first five models each use one individual set of features (to allow a comparison of the usefulness of each feature set in isolation). M6 uses all feature sets except words. Finally, M7 uses all five feature sets. The difference in predictive accuracy between models M6 and M7 is of greatest interest to us, as this comparison quantifies the predictive benefit of using word segmentation on the domain names in addition to the other four feature sets. The same seven models were fit to the data in each of the three experiments.

To measure the effectiveness of our approach, we computed the misclassification rate (MCR) and the area under the ROC curve (AUC) on the test set for each model in each experiment. Our primary outcome variable is the AUC, because it summarizes the tradeoff between the true positive rate and the false positive rate of a classifier, where larger values generally indicate that more favorable tradeoffs are available.

We do not, however, suggest any specific threshold on the predicted probability of maliciousness above which to assign a label of "malicious" to a domain name. The choice of this threshold depends on the relative costs of false positives and false negatives in the particular problem to which this approach is being applied (see [22] for a discussion of this tradeoff). Our MCR results are based on a naive threshold of 0.5.

## VI. RESULTS

Table III contains a summary of the fits of the seven models for the "balanced" data set. For the balanced data, the TLD

|    | Feature Sets | MCR | AUC | # Features | # $\neq 0$ |
|----|---|---|---|---|---|
| M1 | Basics | 0.435 | 0.595 | 22 | 6 |
| M2 | Characters | 0.478 | 0.536 | 36 | 29 |
| M3 | TLD | 0.288 | 0.763 | 489 | 101 |
| M4 | Log-likelihood | 0.492 | 0.512 | 22 | 13 |
| M5 | Words | 0.373 | 0.667 | 24772 | 4588 |
| M6 | M1 + M2 + M3 + M4 | 0.297 | 0.771 | 569 | 104 |
| M7 | M6 + Words | **0.277** | **0.813** | 25341 | 2928 |

TABLE III
SUMMARY OF MODEL FITS TO BALANCED DATA

feature set resulted in the best predictions on the test set among the models that used an *individual* feature set (M1 - M5), based on the MCR and the AUC. M7, which includes all five feature sets, is the best model, as measured by MCR and AUC, and decreases the MCR of M6 (the same as M7 except excluding words as features), by about 2% (on an absolute scale) and increases the AUC by about 4% (on an absolute scale). M7 has slightly fewer than 3000 active (nonzero) features.

Table IV contains a summary of the fits of the seven models for the "unfiltered cellular" data set. The results on this data set are similar to those on the balanced data set, except for two notable differences:

1) The baseline rate of maliciousness is much lower (14.7% compared to 50%), and the improvement in MCR as the models grow more complex is smaller (on the absolute scale and the relative scale) than in the case of balanced data.
2) The best model among M1 - M5 is the model using the Words feature set, rather than the TLD feature set, as was the case in the balanced data set experiment.

As before, M7 provides the best performance across the models, and in this case is substantially better with respect to AUC than M6.

|    | Features | MCR | AUC | # Features | # $\neq 0$ |
|----|---|---|---|---|---|
| M1 | Basics | 0.146 | 0.563 | 22 | 13 |
| M2 | Characters | 0.145 | 0.576 | 36 | 27 |
| M3 | TLD | 0.140 | 0.657 | 479 | 99 |
| M4 | Log-likelihood | 0.146 | 0.542 | 22 | 18 |
| M5 | Words | 0.137 | 0.708 | 77866 | 12568 |
| M6 | M1 + M2 + M3 + M4 | 0.137 | 0.696 | 559 | 158 |
| M7 | M6 + Words | **0.125** | **0.779** | 78425 | 9938 |

TABLE IV
SUMMARY OF MODEL FITS TO UNFILTERED CELLULAR DATA

Table V contains a summary of the fits of the seven models for the "filtered cellular" data set. The results on this data set are similar to those on the "unfiltered cellular" data set, except that the MCR rates decrease at a faster rate, and the AUC is higher.

|    | Features | MCR | AUC | # Features | # $\neq 0$ |
|----|---|---|---|---|---|
| M1 | Basics | 0.240 | 0.593 | 22 | 11 |
| M2 | Characters | 0.242 | 0.597 | 36 | 23 |
| M3 | TLD | 0.211 | 0.681 | 323 | 33 |
| M4 | Log-likelihood | 0.250 | 0.557 | 22 | 9 |
| M5 | Words | 0.212 | 0.720 | 38025 | 8578 |
| M6 | M1 + M2 + M3 + M4 | 0.199 | 0.744 | 403 | 119 |
| M7 | M6 + Words | **0.167** | **0.817** | 38428 | 5416 |

TABLE V
SUMMARY OF MODEL FITS TO FILTERED CELLULAR DATA

## VII. DISCUSSION

We designed the first experiment to be similar to that of [11], in which the authors constructed, among other data sets, a balanced data set with 15,000 benign URLs sampled from DMOZ. Our lowest error rate of approximately 28% is much higher than their lowest error rate of about 1.5%. They, however, use many more features than we do, including lexical information from the path of the URL and various pieces of information gathered from DNS lookups and WHOIS queries. We also use different sources of malicious domains and a different outcome variable, making a direct comparison impossible. The relative improvement in predictive accuracy that we observed by using word segmentation (M7) compared to omitting word segments as features (M6) – 6.7% and 5.4% for MCR and AUC, respectively – suggests to us that our approach could complement their more expensive, but potentially more accurate method.

In the second and third experiments, we found that the MCR decreased by 1.2% and 3.2% on the absolute scale, and by 8.8% and 16.1% percent on the relative scale, respectively, when using M7 instead of M6. Furthermore, the AUC increased by 8.3% and 7.3% on the absolute scale and 11.9% and 9.8% on the relative scale for experiments 2 and 3, respectively, when using M7 instead of M6. These types of gains, demonstrated on a sample of real-world data, would represent substantial improvements if they were to be implemented at scale. Figure 1 shows ROC curves for the seven models fit to the filtered cellular data set (Experiment 3).

We note that although the model with the most *sets* of features, M7, performed the best in each of the three experiments, it wasn't the case that this model contained the most nonzero, or *active*, features. To the contrary, M7 contained fewer active features than M5 in each of the three experiments, because in the presence of the features from the first four sets, not as many individual words were found to be useful (i.e. were active) in predicting maliciousness. This demonstrates the efficacy of lasso-penalized logistic regression for variable selection.

In addition to improving the accuracy of predictions of malicious domains, these logistic regression models provide characters most strongly associated with malicious domains were all rare: "x", "z", "q", "y", and "j". For both the raw and the normalized log-likelihood features, we found that low log-likelihoods (the bottom two deciles, specifically) were predictive of maliciousness, and for unnormalized log-likelihoods, membership in the top decile was associated with benign domains. This suggests that some malicious domains in our data set may have been registered using an unsophisticated algorithm that did not attempt to mimic English language words, but simply combined long strings of numbers and digits. The TLDs that were most strongly associated with malicious domains were ".co", ".us", and ".eu", whereas the safest TLDs were ".de" and ".gov".

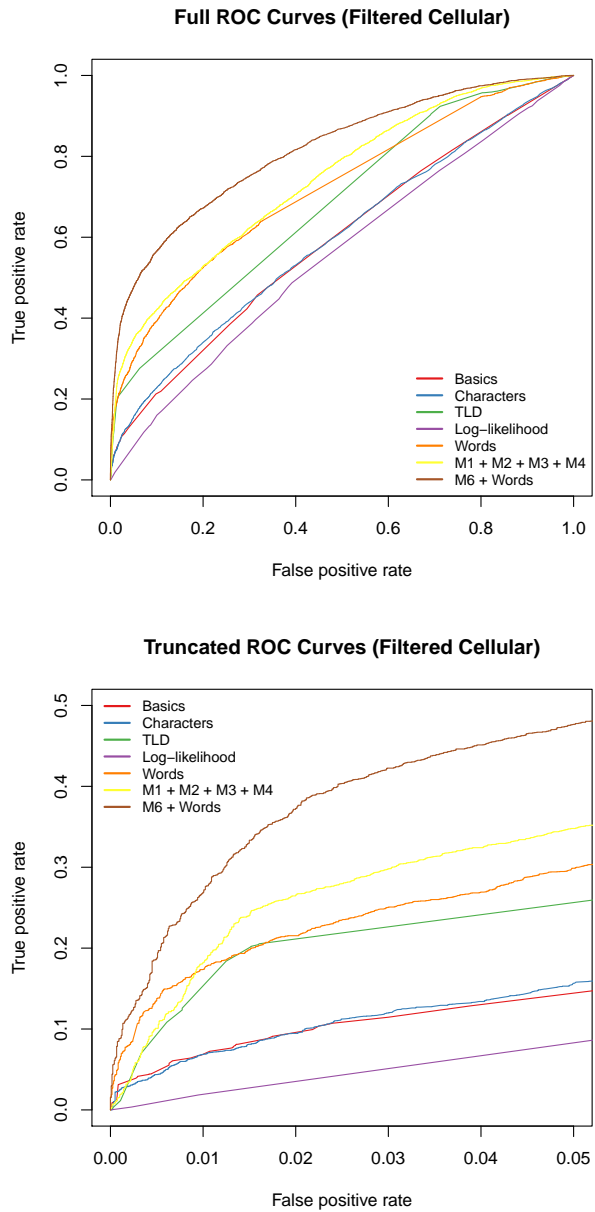

Fig. 1. ROC curves for the filtered cellular data set

Of the 5416 nonzero features in M7 (on the "filtered cellular" data set), 5327 of them were words. Among the largest 400 out of these 5327 coefficients (i.e. those *most* strongly associated with maliciousness) were several words that fell into groups of related words, which we manually labeled in the following list:

1) **Brand names:** rayban, oakley, nike, vuitton, hollister, timberland, tiffany, ugg
2) **Shopping:** dresses, outlet, sale, dress, offer, jackets, watches, deals
3) **Finance:** loan, fee, cash, payday, cheap
4) **Sportswear:** jerseys, kicks, cleats, shoes, sneaker
5) **Basketball Player Names (associated with shoes):** kobe, jordan, jordans, lebron
6) **Medical/Pharmacy:** medic, pills, meds, pill, pharmacy
7) **Adult:** webcams, cams, lover, sex, porno
8) **URL spoof:** com

Several of these groups of words are well-known to be popular in phishing campaigns, such as the brand names [7], shopping-related words, adult-themed words, and words associated with online pharmacies. URL-spoofing words such as "com" (and others such as "www", "login", "mail", and "search", which were among the top 600 most malicious words) were also found to be associated with maliciousness in prior studies [6], [11]. It was especially interesting to be able to discover that certain basketball players' names are associated with malicious domains – theoretically, these names would change over time as new basketball players became popular, and a model like M7, trained on a sliding window of fresh data, would detect the new names.

There were also several interesting words associated with benign domains, which we also found easy to manually group together:

1) **Locations:** european, texas, india, europe, vermont, zealand, washington, colorado
2) **Hospitality Industry:** inn, ranch, motel, country
3) **Common Benign Numbers:** 2000, 411, 911, 2020, 365, 123, 360
4) **Realty:** realty, builders, homes, properties, estate
5) **Small Businesses:** rentals, outfitters, lumber, audio, funeral, flower, taxidermy, inc, golf, law, farm, chamber, farms, rider, photo

interpretable coefficients that describe associations between the features and domain maliciousness. We will interpret the fit of M7 to the "filtered cellular" data set (experiment #3) in detail. Of the 22 basic features, 11 were nonzero. Among them, a domain length of 8-9 characters was predictive of maliciousness, replicating a result found by McGrath and Gupta [7], and additionally domains in the top decile of length ($> 18$ characters) were predictive of maliciousness. Large numbers of hyphens (3 or more) and digits (2 or more) were both found to be positively associated with maliciousness as well; the result pertaining to digits is also consistent with McGrath and Gupta [7]. Among character features, the five

6) **Geographical Features:** creek, hills, lake, ridge, river, valley, springs, grove, mountain, sky, island

Although digits were generally associated with maliciousness, there were certain common numbers, such as 411, 365, and 123 that appear to be used by legitimate, benign domains on a regular basis. Also, the set of words that describe geographical features (creek, hills, etc.) are associated with benign domains, and in our examination, they often appear in benign websites for municipalities, golf courses, realty groups, and rental properties.

## VIII. Conclusion

Given the growing popularity of social media and blogs, as well as the ubiquitousness of smartphones, it is of great interest for users and cellular network providers to be able to proactively identify malicious domains. We find that word segmentation applied to domain names adds substantial predictive power to a logistic regression model that classifies domain names based on their lexical features. In our experiments on real-word data, models that used word segmentation decreased relative misclassification rates and increased relative AUC rates by roughly 10% compared to similar models that didn't use word segmentation. Our method also provides interpretable results that show which words attract users to malicious sites. These words would naturally change over time as attackers change their methods, but a model such as the one presented here could be fit to a sliding window of recent data to continuously monitor and detect new words that are being used in malicious domains.

Our results, of course, depend on the data we used: "fresh" domains visited on a cellular network and WoT reputation ratings as the outcome. Using different thresholds for the outcome variable, different time lags for the query that gathered the WoT reputation ratings, or a different outcome variable altogether (such as one that is specifically tailored to mobile traffic) are interesting directions for future work. We also plan to investigate further into the relative performance of our models on brand-new domains vs. relatively low-traffic domains, since our data set of domains that had not been visited in the prior 30 days consists of a mixture of these two types of domains.

The output of our lightweight method could potentially be used to apply near-real time detection of suspicious domains when a device attempts to visit a domain in a cellular network. If a domain is estimated to have a high probability of being malicious based solely on its name, then a more expensive analysis (such as web content-based analysis) could be used to determine further action, such as blocking the site or inserting a "speed bump". In this way, the word segmentation techniques described here could improve existing systems that use machine learning to detect malicious domains by generating thousands of additional features with which to classify domains.